\documentclass[11pt]{article}
\usepackage{moriond,epsfig}
\usepackage{graphicx,amssymb,amsmath,wasysym,subfigure}

\def\be{\begin{equation}}
\def\ee{\end{equation}}
\def\bea{\begin{eqnarray}}
\def\eea{\end{eqnarray}}

\begin{document}

\begin{flushright}
DESY 11-089
\end{flushright}

\vspace*{4cm}
\title{Neutralino dark matter with a Light Higgs}

\author{ Andreas Goudelis }

\address{Deutsches Elektronen-Synchrotron (DESY), Notkestrasse 85,\\
D-22603 Hamburg, Germany}

\maketitle\abstracts{
We examine the neutralino dark matter (DM) phenomenology in supersymmetric scenarios
with nonuniversal Higgs masses (NUHM) at the gauge coupling unification scale
that can acommodate a light Higgs boson, where the correct relic
density is obtained mostly through the annihilation into a pseudoscalar A. 
Our analysis shows that most
part of the A pole region can produce detectable gamma-ray and antiproton signals. We further 
focus on uncertainties influencing the results in indirect and mainly direct detection.}

\section{Introduction}
\subsection{The model}\label{subsec:TheModel}
One of the major experimental constraints on the constrained-MSSM parameter space comes from the LEP-2
limits on the lightest higgs boson mass. In particular, LEP-2 set a lower bound of $114.4$ GeV
for this mass \cite{hlim}, excluding the largest part of the model's viable parameter space. However,
strictly speaking, this bound only applies to the SM. It comes mostly from
searches in the Higgsstrahlung channel, which in the case of the MSSM is actually not
identical to the SM one. In particular, 
$\sigma_{\mbox{\begin{tiny}MSSM\end{tiny}}}(e^+ e^- \rightarrow hZ) = \sin^2(\beta - \alpha) 
\sigma_{\mbox{\begin{tiny}SM\end{tiny}}}(e^+ e^- \rightarrow hZ)$ \cite{HaberNP,djouadihiggs}.
Hence, the LEP2 bound $m_h \gtrsim 114$ GeV applies to the MSSM only
if $\sin^2(\beta - \alpha) = {\cal{O}}(1)$. This is the case in cMSSM/mSUGRA scenarios.

This limit can actually be partially circumvented once some of the cMSSM constraints
are relaxed. In particular, it has been pointed out \cite{NUHM1,NUHM2,mynusm} that relaxing the requirement for
higgs mass universality at the GUT scale can effectively reduce the $\sin^2(\beta - \alpha)$
factor, leading to a smaller cross-section and thus to weaker bounds on the lightest
higgs mass.
\\
Our particular model \cite{Das:2010kb} is characterised by the
following parameters
\begin{equation}
 m_{1/2}, \ A_0,  \ \mbox{sign}(\mu), \ \tan\beta, \ m_0, \ m_{H_u}^2, \  m_{H_d}^2
\end{equation}
where the GUT-scale common scalar mass $m_0$ concerns all scalars but the two Higgs bosons.

We examined the neutralino dark matter - related phenomenology of this model, notably the behavior of the
relic density over some region of the parameter space, the prospects for indirect
detection as well as the constraints coming from direct DM detection experiments. 
Viable parameter space points have to satisfy a number of constraints:
\\
{{\bf - Higgs boson mass limit:}} 
In the {\it non decoupling} 
region where the $A$ boson
becomes very light the lower limit of $m_h$ goes down to 93 GeV or even lower. 
We consider that the parameter space with $\sin^2(\beta-\alpha) < 0.3$ (or, 
$\sin(\beta-\alpha) \lesssim 0.6$), and $93<m_h<114$ is in agreement with the LEP2 limit \cite{hlim}.
Consequently, the coupling of the heavier Higgs boson to the $Z$ boson
($g_{ZZH} \propto \cos(\beta-\alpha)$)
becomes dominant and this makes the heavier Higgs boson SM - like, so the 
LEP-2 $114$ GeV  limit starts applying for the heavier $CP$-even higgs
boson. On the other hand, in the decoupling region 
$\sin(\beta-\alpha) \sim 1$, which means that the $114$ GeV limit applies
to the lightest higgs. Given the fact that there exists an uncertainty of about
3~GeV in computing the mass of the light Higgs boson \cite{higgsuncertainty}, we
accept a lower limit of $111$~GeV.
\\
{\bf - $Br(b\rightarrow s\gamma)$ constraint:} 
We demand \cite{kim,bsg-recent} $2.77 \times 10^{-4} < Br (b \rightarrow s \gamma) < 4.33 \times 10^{-4}$. 
\\
{\bf - $Br(B_s\rightarrow \mu^+ \mu^-)$ constraint:} 
We further impose the important $Br(B_s \to \mu^+ \mu^-)$ constraint
coming from CDF, \cite{CDF} ${\rm Br} ( B_s \to \mu^+ \mu^-) < 5.8 \times 10^{-8}$ (at ${\rm 95\,\%\,C.L.}$), 
which has recently been improved to $< 4.3 \times 10^{-8}$ 
at $95\%$ C.L \cite{Morello:2009wp}.
\\
{\bf - WMAP constraint :}
In computing the relic density constraint, we consider 
the 3$\sigma$ limit of the WMAP data \cite{WMAPdata} $0.091 < \Omega_{CDM}h^2 < 0.128$. 
Here $\Omega_{CDM}h^2$ is the dark matter relic density in units of the critical
density and $h=0.71\pm0.026$ is the Hubble constant in units of
$100 \ \rm Km \ \rm s^{-1}\ \rm Mpc^{-1}$. We use the 
code micrOMEGAS \cite{micromegas} to compute the neutralino relic density.

\subsection{Dark matter detection}\label{subsec:DMdetection}

There exist two main modes of dark matter detection, usually referred to as ``indirect'' and
``direct''. Indirect detection is based on the principle that if DM can annihilate (or decay), 
in the early universe in order to give the measured relic density, this process should
also occur today throughout the galaxy (and beyond), so we could hope to detect its annihilation
products: gamma-rays, positrons, antiprotons and neutrinos. 
Direct detection of DM relies on the fact that WIMPs may interact with (scatter on) ordinary matter.
This scattering is an in principle measurable effect and indeed a huge effort is currently
being developped worldwide to measure potential signals coming from DM scatterings upon large
underground detectors.

Concerning indirect detecion, in this work we compute the gamma-ray signals 
at intermediate galactic latitudes \cite{Stoehr:2003hf} in the spirit of elliminating
as much as possible uncertainties coming from the DM ``halo profile'' (i.e. its distribution
in the galaxy) as well as background contributions to the spectrum. These gamma-rays
can be detected by the Fermi satelite \cite{Abdo:2010nz}.
For the case of antiprotons, we compute the prospects for detection in the AMS-02 mission \cite{Goy:2006pw}. 
We adopt a semi-analytical treatment
of the diffusion equation \cite{Maurin:2006hy,Maurin:2006ps,Lavalle:1900wn}, presenting results for the
so-called MAX propagation model. 

Finally, we compare the neutralino-nucleon spin-independent scattering cross-section to some
of the tightest bounds available in the literature \cite{cdms-2,xenon100,Kopp:2009qt}. 
As it has been pointed out, uncertainties are not absent
in direct detection as well. These can be of a twofold nature: First of all, when giving exclusion
bounds on the $(m_\chi, \sigma_{\chi (p,n)})$ plane, a set of astrophysical assumptions (as well as
some assumptions on the passage from the nuclear to the nucleonic level) have already been
made. These assumptions are thought to have a small impact on the results, the mangitude
of which depends, among other factors, on the considered mass range 
\cite{McCabe:2010zh,Green:2010ri,Weber:2009pt,Salucci:2010qr,Vogelsberger:2008qb}. 
Secondly, there are often
uncertainties in the cross-section computations performed by theorists in specific models. In our
case, what is of relevance is one of the parameters entering the passage from the partonic to the
the WIMP-nucleon scattering cross-section, denoted by $f_{T_s}$. This parameter is actually related
to the strange quark content of the nucleon. Its value can be either measured or estimated through
lattice QCD methods. The DarkSUSY \cite{darksusy} code, which was used to compute the cross-section, adopts a 
default value of $f_{T_s} = 0.14$. However, recent lattice simulations \cite {Cao:2010ph,lattice} point towards much lower
values, of the order of $0.02$, being even compatible with zero. The effect of this uncertainty
has been quantified \cite{Ellis:2008hf} and is known to range from negligible to very large, depending on the specific
mechanism driving the scattering cross-section. In what follows we shall quantify the effect of this
uncertainty showing that it is really crucial in assessing the viability of our models.

\section{Results}
\subsection{Indirect detection}\label{subsec:indirect}

We performed \cite{Das:2010kb} two scans in the model's parameter space:\\
- In the first one, we fix $\tan\beta(=10)$,  $m_0=600~$ GeV, $A_0=-1100$ GeV, $\mbox{sign}(\mu)>0$. 
Then, we vary the mass parameters $m_{h_u}$ ($0<m_{h_u}<m_0^2$) and $m_{h_d}$ 
($-1.5 m_0^2<m_{h_d}<-0.5m_0^2$) to obtain light neutralino dark matter 
consistent with light Higgs masses ($m_{H,A}\le 250~$GeV) at the electroweak scale.
$\mu$ and $m_A$ are derived quantities. We note that the high $\mu$ parameter values
obtained in this scenario correspond to an essentially pure bino LSP.\\
- In our second scan, we fix $m_0$ at the very similar value 
$m_0=600~$GeV, while $A_0=-1000~$GeV is chosen to make $b \rightarrow s \gamma$ less restrictive.  
We set  $m_{h_u}$ ($=2.4m_0^2$) and vary $m_{h_d}$ ($-0.3 m_0^2<m_{h_d}< 0.1$) 
with $m_{1/2}$ to obtain the WMAP-compatible regions for neutralinos.
Our scan renders very small $\mu$ values ($150<\mu<300$), 
consequently the LSP can have large Higgsino components.\\
In both cases, we look for points satisfying
the relic abundance requirement while passing all constraints previously mentioned.
The viable points are then scattered on the $m_{1/2}-m_A$ plane as red dots.

\begin{figure}
 \begin{center}
  \epsfig{file=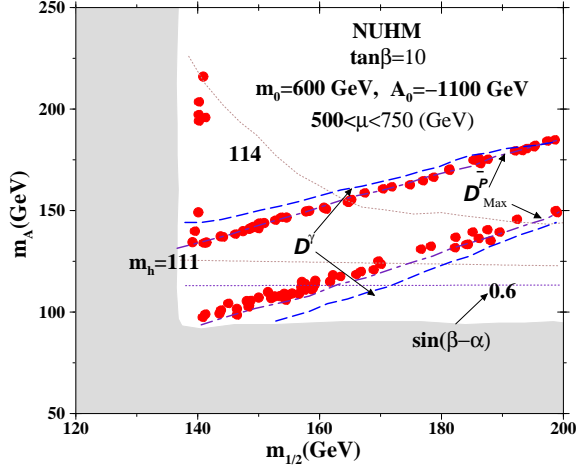,height=2.5in}
  \caption{
Viable points in the $m_{1/2}-m_A$ plane.
Neutralino masses of $\sim 55-65$ GeV 
correspond to the {\it Light Higgs Boson} region. Detectability of the
photon and anti-proton signals are represented by $D^{\gamma}$ and $D^{\bar p}_{Max}$
lines.} \label{fig:Light}
 \end{center}
\end{figure}

\begin{figure}
 \begin{center}
  \epsfig{file=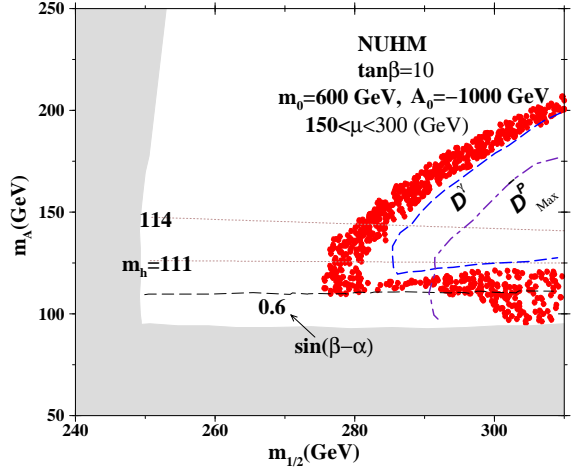,height=2.5in}
  \caption{
Same as Fig.\ref{fig:Light}, except that the {\it light Higgs boson} zone
is shifted to larger neutralino mass values.} \label{fig:Heavy}
 \end{center}
\end{figure}

Our results can be seen in figures \ref{fig:Light} and \ref{fig:Heavy}. Appart from the
WMAP-compliant points, we demonstrate in the same plots regions excluded by other constraints
(gray regions), higgs mass isocontours, as well the parameter space regions that can
be probed during a $3$-year data acquisition period for Fermi and AMS-02: all points lying 
inside the two parallel lines in the case of fig.\ref{fig:Light} and on the left or below the
lines in fig.\ref{fig:Heavy} can -in principle- be probed.

In the first set of scenarios, there are mainly two mechanisms that can generate the
correct relic density: quasi-resonnant annihilation through a $A$ or $H$ pole,
extending along the direction of the line where  $2 m_{\chi_1^0} \approx m_A, m_H$,
or the light Higgs pole at low $m_{1/2}$ and along the $m_A$ direction.

In the second set of scenarios the neutralino self-annihilation cross-section is enhanced
kinematically as before (i.e. we are once again sitting near the $A$ pole), but moreover the
neutralino acquires a non-negligible higgsino component which enhances its couplings to the 
higgs bosons.

In both scenarios, we see that the detection prospects are quite good. Significant 
portions of the viable parameter space can be probed. 

In the first case, perspectives are
actually good along the $A$-pole. On the contrary, we see that the light higgs pole seems
to be completely invisible in both channels. This is due to the fact that whereas resonnant
annihilation is an efficient mechanism in the early universe, the cross-section for this 
process tends to zero as the neutralino velocity does so \cite{dmreview,Bernal:2009jc}. 
This is actually the case at present
times, which are of relevance for indirect detection. On the other hand, annihilation through
a pseudoscalar is not that sensitive to changes in the WIMP velocity \cite{dmreview}, so the cross-section 
remains relatively high even at present times.

In the second set of scenarios, the prospects are actually even better. On the one hand, any 
interference of CP-even higgs bosons is negligible. Moreover, the fact that in this case
the neutralino has a significant higgsino component enhances its couplings to the higgs sector,
an effect which is practically insensitive to velocity changes. So, the cross-section remains
quite stable at present times.

\subsection{Direct detection and associated uncertainties}\label{subsec:direct}
\begin{figure}
 \begin{center}
  \epsfig{file=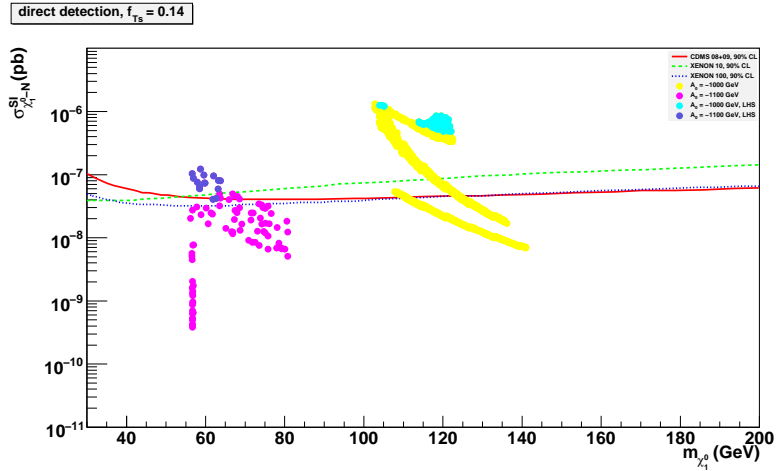,height=4in,angle=-90}
  \caption{
$(m_{\chi^0_1}, \sigma_{\chi^0_1-N}^{SI})$ combinations 
along with the relevant exclusion limits from direct detection
experiments for $f_{Ts}^{(p, (n))} = 0.14$. 
Points lying above the lines are excluded according to the published
limits. The light blue and  the dark blue points represent the
{\it light Higgs boson} regime for the two scenarios at stake.} \label{fig:fTs014}
 \end{center}
\end{figure}

\begin{figure}
 \begin{center}
  \epsfig{file=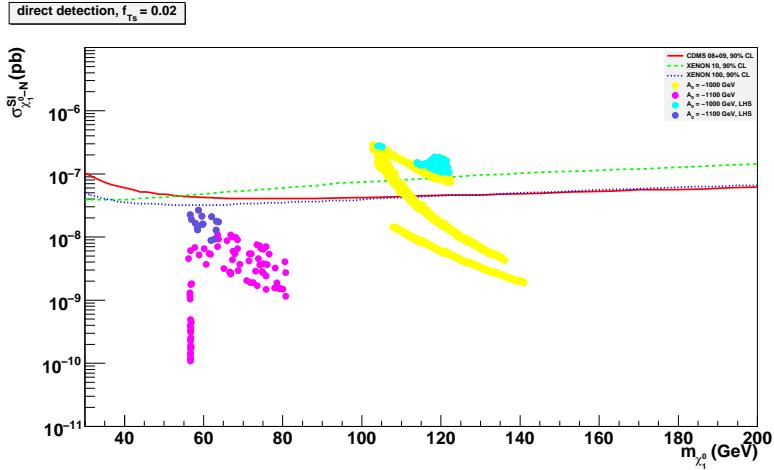,height=4in,angle=-90}
  \caption{
As in fig.\ref{fig:fTs014} but for $f_{Ts}^{(p, (n))} = 0.02$.} \label{fig:fTs002}
 \end{center}
\end{figure}

As a final step, we computed the spin-independent neutralino-nucleon scattering cross-section.
The results can be seen if figures \ref{fig:fTs014} and \ref{fig:fTs002}, for two different
values of the $f_{T_s}$ parameter, namely the default DarkSUSY value of $0.14$ and a reduced
value of $0.02$ respectively.

From fig.\ref{fig:fTs014} we see that our NUHM model seems to be hopelessly
excluded, especially when it comes to the Light Higgs Scenarios. The situation
seems to be however quite different once we look at fig.\ref{fig:fTs002}.

A comparison of the two figures demonstrates the importance of the uncertainty in the value
of $f_{T_s}$. We see that once we reduce its value from the default one towards
lower values the cross-section is reduced by something like an order of magnitude.
We further remind that $f_{T_s}$ estimates are even compatible with zero. 
Moreover, once a set of astrophysical or nuclear uncertainties is taken into account (local density, 
velocity distribution, nuclear form factors), the experimental limits may be considered to bare
an additional uncertainty of a factor $~3 - 4$. We thus see that stating whether a model is excluded
by direct detection data or not may be a tricky issue. All potential sources of uncertainty
should be examined before definitively ruling out a model. Fortunately, a large effort is
being devoted by several groups in quantifying these uncertainties and incorporating them
in a systematic manner, both in the limits published by experimental collaborations and in
the calculations performed by theorists.

\section*{Acknowledgments}
The work of A.G. is supported in part by the Landes-Exzellenzinitiative Hamburg. A.G.
would further like to thank the organisers of the Rencontres de Moriond 2011 for 
their warm hospitality.

\end{document}